\documentclass[twocolumn,prl,showpacs,preprintnumbers,amsmath,amssymb]{revtex4}
\usepackage{graphicx}
\usepackage{dcolumn}
\usepackage{bm}
\begin{document}

\title {Interstitial-Boron Solution Strengthened WB$_{3+x}$}

\author{Xiyue Cheng$^1$}
\author{Wei Zhang$^{1,2}$}
\author{Xing-Qiu Chen$^1$}
\email[Corresponding author:]{xingqiu.chen@imr.ac.cn}
\author{Haiyang Niu$^1$}
\author{Peitao Liu$^1$}
\author{Kui Du$^{1,2}$}
\email[Corresponding author:]{kuidu@imr.ac.cn}
\author{Gang Liu$^1$}
\author{Dianzhong Li$^1$}
\author{Hui-Ming Cheng$^1$}
\author{Hengqiang Ye$^1$}
\author{Yiyi Li$^1$}

\affiliation{$^1$ Shenyang National Laboratory for Materials
Science, Institute of Metal Research, Chinese Academy of Sciences,
Shenyang 110016, China}

\affiliation{$^2$ Beijing National Center for Electron Microscopy,
Tsinghua University, Beijing 100084, China }

\date{\today}

\begin{abstract}

By means of variable-composition evolutionary algorithm coupled with
density functional theory and in combination with
aberration-corrected high-resolution transmission electron
microscopy experiments, we have studied and characterized the
composition, structure and hardness properties of WB$_{3+x}$ ($x$ $<$
0.5). We provide robust evidence for the occurrence of
stoichiometric WB$_3$ and non-stoichiometric WB$_{3+x}$
both crystallizing in the metastable $hP$16 ($P$6$_3$/$mmc$) structure.
No signs for the formation of the highly debated WB$_4$
(both $hP$20 and $hP$10) phases were
found. Our results rationalize the seemingly contradictory
high-pressure experimental findings and suggest that the
interstitial boron atom is located in the tungsten layer and
vertically interconnect with four boron atoms, thus forming a
typical three-center boron net with the upper and lower boron layers
in a three-dimensional covalent network, which thereby strengthen
the hardness.
\end{abstract}


\maketitle

Typical ultrahard or superhard materials
\cite{Kaner,Brazhkin2,Gilman} (i.e., diamond,
\emph{c}-B$_2$CN, \emph{c}-BN,
$\gamma$-B$_{28}$ \cite{Oganov,Zarechnaya} and most
recently synthesized nanotwinned \emph{c}-BN \cite{Tian}) would
require three-dimensional (3D) bonding networks commonly
consisting of high densities of strong covalent bonds, atomic
constituents and valence electrons as well as nano-scale grains.
Currently, the most powerful way to yield the high densities of
these factors is the synthesis under high pressure conditions.
However, in recent years transition metal borides
(OsB$_2$, ReB$_2$, CrB$_4$, and FeB$_4$, etc
\cite{ReB2-1,ReB2-2,ReB2-3,review, Ivanovskii,FeB4, Niu, Gou, Curtarolo})
have attracted extensive interests because of superior
mechanical properties and ambient-condition synthesis without the
need of high pressure, although their hardness
is not as hard as superhardness.

Among those borides, the W-B system has attracted particular
attention since the report \cite{Gu} on WB$_4$ with a measured
superhardnees, $H_v$ of about 46.3 GPa, under the loading force of
0.49 N, as the highest measured hardness among those borides
mentioned above. Although the superhardness of WB$_4$ has been again
confirmed experimentally \cite{Reza} and interpreted theoretically
\cite{Wang} based on the widely accepted $hP$20-WB$_4$
structure \cite{Romans,Nowotny,Reza,Gu}, several subsequent
first-principles calculations denied its existence \cite{Liang,Zhangr,Liquan}.
This structure is neither thermodynamically \cite{Liang}
nor dynamically \cite{Zhangr} stable,
and also not superhard \cite{Liang2} ($H_v$ = 6.5 GPa
according to a recently proposed hardness model
\cite{Chen,22-1}). Instead, those first-principles calculations
\cite{Liang,Zhangr,Liquan} suggested
that the experimentally attributed
WB$_4$ \cite{Reza, Xie, Reza2} should be characterized as a
$hP$16-WB$_3$ phase.

Given the fact that the theoretically proposed
$hP$16-WB$_3$ \cite{Liang,Zhangr,Liquan} is
thermodynamically and mechanically stable,
and its {\small XRD} pattern matches well the experimentally observed
ones \cite{Romans,Reza}, there seems no reason to
suspect the reliability of its composition and structure.
However, it is highly surprising that four recent high-pressure
experimental findings of this phase
\cite{Gu,Xie,Xiong,Liu} yielded conflicting tendency of the pressure
dependence normalized $c$/$a$ ratio.
The more striking fact is that none of them
agrees with the theoretically derived
pressure-dependent $c$/$a$ ratio \cite{Zang} of $hP$16-WB$_3$.
Therefore, this tungsten boride still needs further clarification.

Within this context, by combining first-principles
calculations \cite{VASP1,VASP2} (unless otherwise mentioned,
all calculations have been performed with the
Perdew-Burke-Ernzerh generalized
gradient approximation (GGA-type PBE) \cite{PBE}), variable-composition
evolutionary algorithm search as implemented
recently in {\small USPEX} \cite{USPEX1,USPEX2} and the
aberration-corrected images of high resolution transmission electron
microscopy ({\small Ac-HRTEM}) (method details refers to Supporting information),
we have confirmed the existence of WB$_{3+x}$ ($x$ $<$ 0.5) (including the
stoichiometric $hP$16-WB$_3$) and denied the formation of
the extensively debated boride of $hP$20-WB$_4$ \cite{Reza,Xie,Reza2}.
The previously experimentally attributed WB$_4$
is indeed WB$_{3+x}$. The results uncovered that the structure
can be regarded as a defective $hP$16-WB$_3$
one but with a certain proportion $x$ of extra interstitial boron
locating in the tungsten atomic layers.
Varying $x$, the normalized $c$/$a$ ratio rationalized four puzzling
high-pressure experimental findings \cite{Gu,Xie,Xiong,Liu}.
Importantly, our results revealed that the interstitial boron
solution highlights an effect of strengthening (hardening) to the
mechanical property of WB$_{3+x}$ because of the appearance of the 3D
covalent framework induced by boron solution from the ideal 2D boron
sheets in $hP$16-WB$_3$.

\begin{figure}[htbp]
\begin{center}
\includegraphics[width=0.48\textwidth]{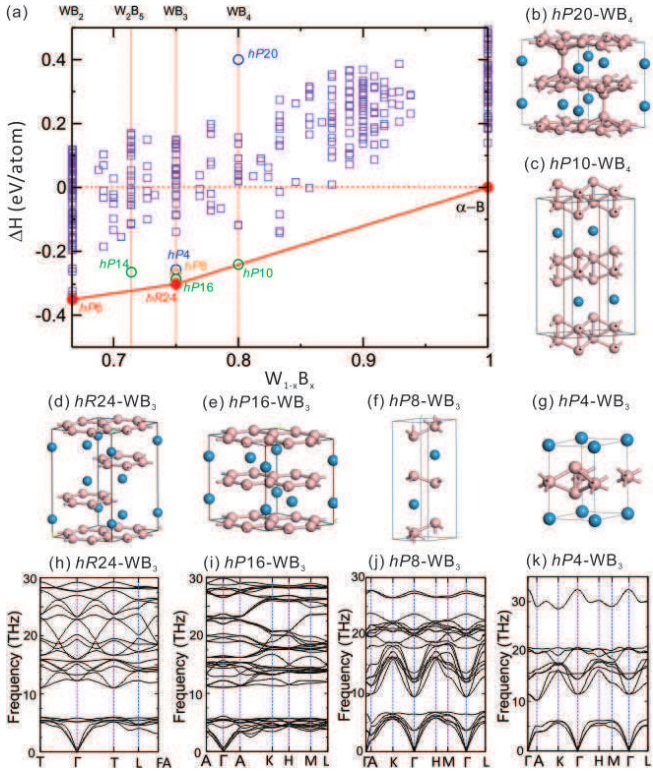}
\end{center}
\caption{(color online) (a) The derived {\small GGA}-type
{\small PBE} enthalpies of formation
predicted by variable-composition evolutionary computations for the
WB$_2$-B system (more details refer to Supporting information).
Every square represents an individual structure and the
most stable ground state phases (solid circles) are connected to a
convex hull. Hollow circles denote metastable phases above the
convex hull.
In panels (b) and (c),the crystal structures of WB$_4$: (b)
the previously experimental attributed structure ($hP$20) and
(c) the {\small USPEX} searched ground state phase ($hP$10);
In panels (d to k), the {\small USPEX} searched crystal structures
and their corresponding phonon dispersions of WB$_3$:
(d,h) $hR$24, (e,i)$hP$16, (f,j) $hP$8, and (g,k) $hP$4.
\label{fig1}}
\end{figure}

\begin{figure}[htbp]
\begin{center}
\includegraphics[width=0.45\textwidth]{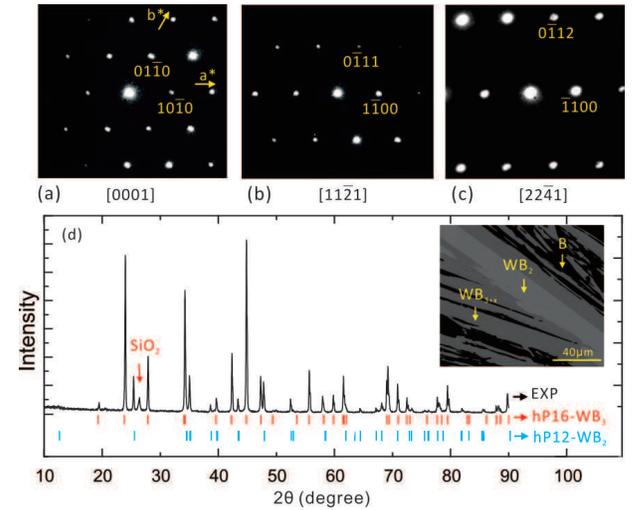}
\end{center}
\caption{(color online)
(a-c) Experimental electron diffraction (ED)
patterns along the [10$\overline{1}$0], [11$\overline{2}$1]and
[22$\overline{4}$1] directions, respectively. (d) The experimental
X-ray diffraction data; reflections of the ideal $hP$16-WB$_3$ and
$hP$12-WB$_2$
are indicated by vertical bars. Right inset
in (d) shows an SEM image of annealed sample. Note that 
amorphous boron could not be distinguished by the XRD pattern.
Besides, the 26.4$\,^{\circ}$ peak origins from SiO$_2$ which was ground by the
much harder WB$_{3+x}$ in the preparation of the x-ray diffraction powder
using an agate mortar and pestle.
\label{fig2}}
\end{figure}

\begin{figure*}[htbp]
\centering
\begin{minipage}[c]{0.785\linewidth}
\includegraphics[width=0.97\textwidth]{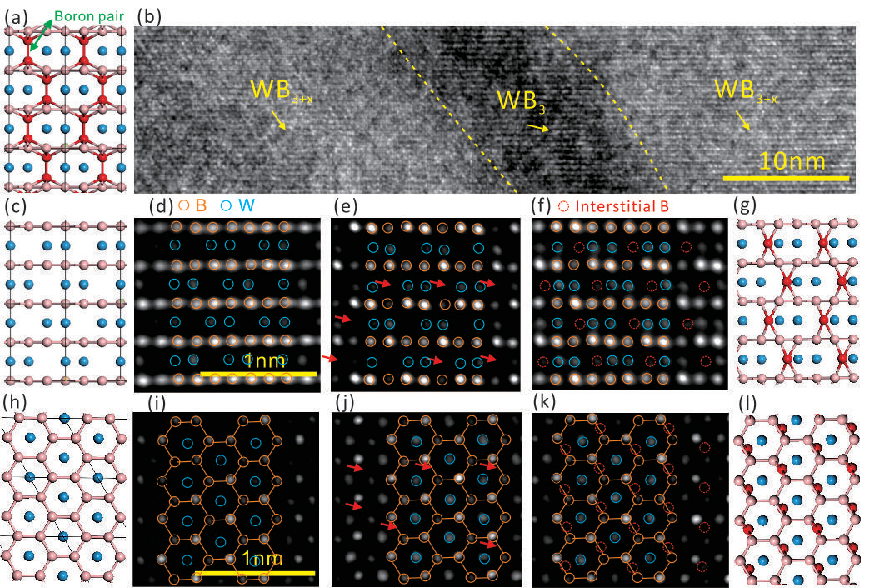}
\end{minipage}%
\begin{minipage}[c]{0.215\linewidth}
\caption{(color online) The {\small Ac-HRTEM} characterization of
the superhard WB$_{3+x}$. (a) The projection along the
[11$\overline{2}$0] direction of the $hP$20-WB$_4$
\cite{Gu, Reza, Xie, Reza2}. (b) The {\small Ac-HRTEM} morphology
with a 10 nm dimension. (c) and (h) The projections along the
[11$\overline{2}$0] and [0001] directions of $hP$16-WB$_3$,
respectively. (d,e,f) and (i,j,k) The {\small Ac-HRTEM} images with
the 1 nm dimension projected along the [11$\overline{2}$0] and
[0001] directions of defective WB$_{3+x}$, respectively. (g and l)
the projections of the structural model along the
[11$\overline{2}$0] and [0001] directions of defective WB$_{3+x}$,
respectively.}
\end{minipage}%
\label{fig3}
\end{figure*}

The {\small USPEX} searched results
are compiled in Fig. \ref{fig1}
(details refer to
Supporting information and Figure S1).
Although the {\small USPEX} found a
lowest-enthalpy $hP$10-WB$_4$
(MoB$_4$-type, Fig. \ref{fig1}(c))
in accordance with four recently
theoretical results \cite{Gou,Liquan,Zhangmgc,Liang3},
its enthalpy
of formation lies
about 2.3 meV/atom above
the $hR$24-WB$_3$ $\leftrightarrow$ $\alpha$-B tieline of the convex hull,
suggesting its instability at $T$ = 0 K. In addition,
the experimentally attributed $hP$20-WB$_4$ phase (Fig. \ref{fig1}(b))
is confirmed to be definitely unstable because its enthalpy is positive,
much higher than the convex hull.
Therefore, we excluded the existence of WB$_4$ at the ground state.
At WB$_3$ composition, the {\small USPEX} searches
demonstrated that the stability of WB$_3$ is
highly robust at the ground state (Fig. \ref{fig1}(a)).
The results suggest that WB$_3$ has a
lowest-enthalpy $hR$24 (\emph{R}$\overline{3}$\emph{m})
structure agreeing with the recent
results \cite{Liang3,Liquan} and has
three metastable phases (Fig. \ref{fig1}(d - g)),
$hP$16 ($P6_3/mmc$)\cite{Liang,Zhangr,Liquan},
$hP$8 ($P\overline{3}m1$), and
$hP$4 ($P\overline{6}m2$) with little energy deviations.
Their optimized lattice parameters are compiled in supporting information,
Table S3.
These four lattices have been confirmed dynamically stable
by the derived phonon dispersions without any imaginary frequencies,
as illustrated in Fig. \ref{fig1}(j - m).
From the phonon densities of states, we further derived the
temperature-dependent Gibbs free energies by also including
zero-point energies and static DFT energies among
these phases (Fig. \ref{fig1}(c)),
revealing a phase transition from the ground state
$hR$24-WB$_3$ to the metastable $hP$16-WB$_3$
phase above 659 K. In order to elucidate the impact of
the exchange-correlation functional on phase transition, we further
employed the localized density approximation (LDA)
potential \cite{LDA} and found consistently
this phase transition above 678 K (Supporting information, Figure S2).

To clarify these theoretical results, we further measured
{\small XRD} patterns of the powder sample annealed for 144
hours at 1523 K (details refer to Supporting information).
As illustrated in Fig. \ref{fig2}(d),
the {\small XRD} pattern
uncovers the mixture of WB$_2$ and WB$_{3+x}$.
Furthermore, from the {\small SEM} image (see,
insert of Fig. 2d), the dendrite of WB$_2$ and WB$_{3+x}$ can be
identified,
whereas the dark contrasting part is amorphous boron
which can not be detectable by XRD.
It is clear that the {\small XRD} pattern of WB$_2$ indicates a
$hP$12 (WB$_2$-type) phase,
in good agreement with the previous
experimental characterization \cite{ReB2-3}.
The {\small XRD}
pattern of WB$_{3+x}$ has been found in accordance with the reported
ones \cite{Reza, Xie, Reza2}
and is exactly same with the
theoretically proposed metastable $hP$16-WB$_3$ phase, rather than
its ground state $hR$24-WB$_3$ phase.
This fact can be interpreted well,
since our annealed temperature of 1523 K is much
higher than our estimated temperature
(659 K (GGA) and 678 K (LDA)) of phase transition (Fig. \ref{fig1}(c)).
Furthermore, the electron diffraction {\small ED} images (Fig.
2(a-c)) and the {\small XRD} pattern (Fig. 2d)
evidence a hexagonal structure of WB$_{3+x}$ with the lattice
parameters of \emph{a} = 5.2055 \AA\,, \emph{c}= 6.3348 \AA\, and
the axial ratio $c/a$= 1.2169, agreeing well with the previously
reported data \cite{Reza, Xie, Reza2}. Moreover, the Vickers
hardness of the polycrystal sample of WB$_{3+x}$ was measured to be
36.7 GPa under a loading force of 1N, which is comparable to the
reported values (31.8 GPa under 1.2 N \cite{Gu} and 38.3 GPa under
1N \cite{Reza}).

\begin{figure}[htbp]
\begin{center}
\includegraphics[width=0.48\textwidth]{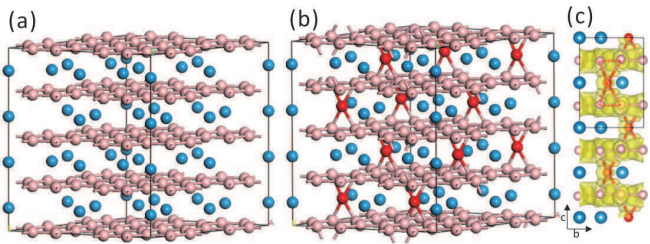}
\end{center}
\caption{(color online) The 2$\times$2$\times$2 supercell of (a)
stoichiometric $hP$16-WB$_3$, (b) non-stoichiometric WB$_{3+x}$, and (c)
the partial isosurface of the ELF (with an isovalue of 0.65)
projected along the [11$\overline{2}$0] of WB$_{3+x}$, mainly
illustrating the three-center covalent boron net as marked by the
dashed line. \label{fig4}}
\end{figure}

\begin{figure}[htbp]
\begin{center}
\includegraphics[width=0.48\textwidth]{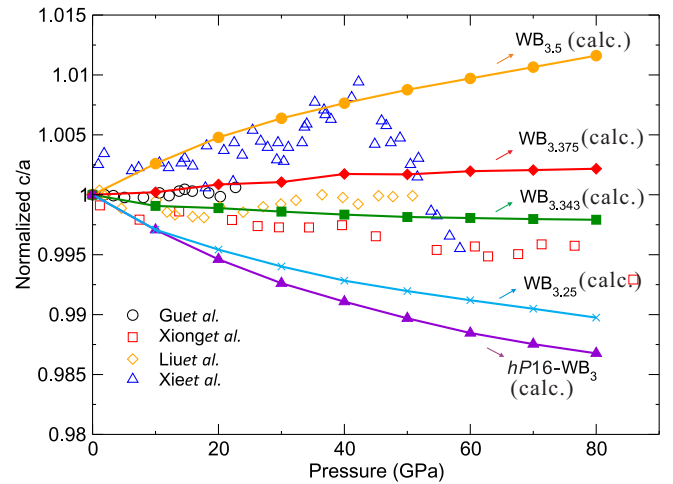}
\end{center}
\caption{(color online) The calculated pressure dependence of the
normalized $c$/$a$ ratio of WB$_{3+x}$ (0$<$x$<$0.5), interpreting
well the seemingly contradictory experimental results
\cite{Gu,Xie,Xiong,Liu}. \label{fig5}}
\end{figure}

Because boron is a weak electronic scatterer, it is impossible
to refine the accurate structure of WB$_{3+x}$ from
{\small XRD} patterns of powder samples.
However, the
{\small Ac-HRTEM} image \cite{Jiaachrtem} provides a powerful tool
to directly visualize the light mass elements (i.e., oxygen and
boron) with the minimum resolution length of about 0.8 \AA\,
(Supporting information). Figure 3b shows the {\small Ac-HRTEM} morphology
of WB$_{3+x}$ and boundaries between the dark
and bright regions can be clearly identified as marked by the dashed
curves.
The {\small Ac-HRTEM images} with the 1nm dimension of Fig. 3(d,e,f)
were selected along the [11$\overline{2}$0]
projection, while Fig. 3(i,j,k) corresponds to the same region
along the [0001] projection. Interestingly, Fig. 3d and 3i perfectly
match the theoretical projections (Fig. 3c and 3h) along the same
directions of the $hP$16-WB$_3$, confirming the existence of
the stoichiometric $hP$16-WB$_3$. For the [11$\overline{2}$0] projection of
$hP$16-WB$_3$, between any two dense boron lines there exists a
tungsten-atom line, which consists of a repeated unit of every two
tungsten atoms separated by a void. However, what out of our
expectation is that the voids are partially occupied in the
WB$_3$/WB$_{3+x}$ boundary (see arrows in Fig. 3e) and fully
occupied in Fig. 3f (see dashed hollow circles). We further identify
that these voids should be occupied by extra boron atoms, rather
than tungsten atoms. If the tungsten atom occupies these voids, the extra
peaks would be expected to appear from the XRD patterns in Fig. 2d.
As further evidence, the extra atoms can be also found in the
{\small Ac-HRTEM} images in Fig. 3j and 3k projected along the
[0001] direction. The fact reflects well the occurrence of the new
composition of WB$_{3+x}$, which shares, accordingly, the same
hexagonal structure with the $hP$16-WB$_3$. In other words,
WB$_{3+x}$ can be considered as defective $hP$16-WB$_3$ in which the extra
$x$ boron atoms occupy the interstitial sites in the tungsten
layers. Moreover, the previously experimentally attributed
$hP$20-WB$_4$ \cite{Reza, Xie, Reza2} have been absolutely excluded,
because no any boron pairs (as
illustrated in Fig. 3a projected along its [11$\overline{2}$0]
direction) can be identified in the {\small Ac-HRTEM} images (Fig.
3(d-f)).

We further performed a series of first-principles computational
experiments to analyze the exact position of the
interstitial boron atoms within a 2$\times$2$\times$2
supercell of $hP$16-WB$_3$ (Fig. 4a). According to the
{\small Ac-HRTEM} images, the interstitial boron atoms are
placed in the same layers of tungsten atoms but deviated from
all tungsten atoms if viewed from the [0001] direction.
After the relaxation, these interstitial boron atoms
still locate in the layers of tungsten atoms
but now each interstitial boron binds four nearest
boron atoms, equivalently, from its upper and lower boron hexagonal
rings, as shown in Fig. 4b.
These positions nicely agree
with our {\small Ac-HRTEM} images (Fig. 3f and 3k).
In fact,
there exists sixteen equivalent interstitial sites in this
supercell, determining that the maximum \emph{x} would be 0.5. To keep
the full agreement with the experimental {\small Ac-HRTEM} results
in Fig. 3f and 3k, at least eight interstitial sites (namely,
WB$_{3.25}$) have to be occupied by boron. Therefore, the \emph{x}
content in the boundary region in Fig. 3e and 3j should be below 0.25.
Figure 3g and 3l illustrate the projections along the
[11$\overline{2}$0] and [0001] directions for WB$_{3.25}$, in
agreement with the {\small Ac-HRTEM} images (e.g., Fig 3f and 3k)).
In addition, it can be seen that the {\small Ac-HRTEM} images of WB$_{3+x}$
along [11$\overline{2}$0] and [0001] directions are well consistent
with the simulated images (Supporting information, Figure S3) of the proposed
structural model.

Another compelling support to the defective WB$_{3+x}$ is the
normalized pressure dependent $c$/$a$ ratio (Fig. 5). In agreement
with the reported results by Zang \emph{et. al.} \cite{Zang}, the
stoichiometric $hP$16-WB$_3$ exhibits a large negative pressure dependence.
However, with increasing $x$, the pressure dependence is
substantially elevated. As a result, this behavior interprets well
the apparently contradictory results from four recent high pressure
measurements. It can be seen that with \emph{x} = 0.343 the
theoretical trend is similar to that of the experimental
observations \cite{Xiong,Liu} whereas with $x$ = 0.375 the normalized
$c$/$a$ ratio remains nearly unchanged in
the pressure region considered here, which again
agrees well with another experimental findings \cite{Gu}.
When \emph{x} reaches its maximum of 0.5, the normalized
\emph{c}/\emph{a} ratio rises significantly from 1 to 1.0115 with
increasing pressure up to 45 GPa, matching the experimental
observation by Xie $et$ $al$ \cite{Xie}, although our theoretical
trend does not reproduce the experimentally observed quick drop by
less than 1 from 45 GPa to 60 GPa.

Mechanically, upon different conditions (temperature and pressure)
of synthesis a certain proportion of interstitial boron atom
diffuses in the space featured by any two ideal 2D boron sheets
(Fig. 4a) in $hP$16-WB$_{3}$. This kind of boron solution contributes an
extrinsic component to the superhardness and strengths of WB$_{3+x}$
due to the formation of 3D covalent network through the connection
of the three-center covalent boron nets as highlighted by dashed
lines in Fig. 4c. Our results demonstrated that the interstitial
boron atoms in a solid solution way are considered as an efficient
routine to tune the mechanical properties (hardness) of
transition-metallic borides. The findings highlight a promising
factor utilizing the concept of solid solution \cite{Reza2}
to design superhard materials, besides our widely recognized
manipulations of covalent bonds, valence electrons and atomic
constituents as well as nano-scale grains. In addition, despite
of a good level of maturity of structural searches for materials
discovery, those successful methods ({\small USPEX}
\cite{USPEX1,USPEX2}, {\small AIRSS} \cite{AIRSS},
{\small MAISE} \cite{MAISE1}, {\small  CALYPSO} \cite{Liquan}, and
{\small AFLOW} \cite{Curtarolo}, etc)
may need further algorithm
implementations on the search of this class of defective or
non-stoichiometric structures. Finally, our detailed theoretical and
experimental studies of this tungsten boride demonstrate
that the compositions and structures of many reportedly known
transition metal borides need to be further investigated in-depth
using advanced techniques coupled with the art-of-the-state
first-principles calculations, in terms of the characterization
difficulties of boron atoms.

\textbf{Acknowledgements} We are grateful for the useful discussions
with Artem Oganov and Qiang Zhu and for the experimental synthesis
with Jiaqi Wang and Shi Liu. This work was supported by the
``Hundred Talents Project'' of the Chinese Academy of Sciences and
from NSFC of China (Grand Numbers: 51074151, 51174188, 51171188) as
well as Beijing Supercomputing Center of CAS (including its Shenyang
branch) and Vienna Scientific Clusters. This work made use of the
resources of the Beijing National Center for Electron Microscopy.

\end{document}